# Routing Protocols for Mobile Sensor Networks: A Comparative Study

Shahzad Ali[1,2], Sajjad A. Madani[2], Atta ur Rehman Khan[3], Imran Ali Khan[2]

[1]Institute IMDEA Networks, Avenida del Mar Mediterraneo, Madrid, Spain.
[2]COMSATS Institute of Information Technology, Abbottabad, Pakistan.
[3]FSKTM, University of Malaya, Kuala Lumpur, Malaysia.

Email: shahzadali@ciit.net.pk, madani@ciit.net.pk, attaurrehman@siswa.um.edu.my, imran@ciit.net.pk

*Abstract*

*This paper presents a comparison of cluster-based position and non position-based routing protocols for mobile wireless sensor networks to outline design considerations of protocols for mobile environments. The selected protocols are compared on the basis of multiple parameters, which include packet delivery ratio, packet loss, network lifetime, and control overhead using variable number of nodes and speeds. The extensive simulation and analysis of results show that position-based routing protocols incur less packet loss as compared to the non position based protocols. However, position-based protocols require localization mechanism or a GPS for the location information, which consumes energy and affects the network lifetime. Alternatively, non position-based protocols are more energy efficient and provide extended network lifetime.*

## 1. INTRODUCTION

Wireless sensor network (WSN) is a network of sensor nodes that are distributed over a certain area to monitor a physical phenomenon, such as temperature, humidity, or fire. WSN consist of a large number of sensor nodes and a resource rich sink node(s) that acts as a gateway to other networks or a final destination [1]. The sensor nodes are characterized by limited battery power, processing capacity, and memory resources [2]. Therefore, WSNs require low footprint communication schemes, which utilize minimum resources without compromising the required quality of service. Moreover, due to limited energy of the sensor nodes, the energy efficiency is one of the most important design considerations of the WSN routing protocols [3]. However, mobility of sensor nodes makes the routing process challenging and complex, which hampers the energy efficiency of the routing protocols.

An overwhelming majority of the current research on sensor networks consider static networks [1, 2, and 4]. However, there exist many applications [6, 7] which require mobile nodes, for instance, habitat monitoring, battlefield surveillance, and object tracking. In mobile WSNs, unpredictable topology changes and frequent path failures make the routing challenging [2], because path breakage leads to increase in the end-to-end delay and packet loss [8]. Some routing schemes assume that the sensor nodes can directly communicate with the sink node [7, 8, and 9]. This assumption restricts the geographic scalability, but is countered by multi-hop routing schemes. Considering the mobility and resource constrained environment, WSNs require energy aware routing protocols, capable of decreasing the packet loss, and enhancing robustness against mobility of nodes. Furthermore, WSN protocols require scalability and aptitude to coup with dynamicity of the network that is caused by the mobility [10, 11]. To achieve this, backup strategies are required which enable data packets to reach a destination in the presence of mobility [12].

In this study, we focus the comparison of cluster based position and non position-based hierarchal routing schemes based on multiple performance factors, such as energy efficiency, overhead, and packet loss. We select six routing protocols for comparison, which are studied thoroughly for variable network densities and node speeds. The performance metrics that are used for the evaluation of the protocols include, average packet delivery ratio, network lifetime, average number of control packets sent during the protocol operation, and average percentage of packet loss.

The remainder of this paper is organized as follows. Section 2 contains the related work. Section 3 presents the comparison strategy. Section 4 consists of results and discussion, whereas Section 5 concludes the paper.

## 2. RELATED WORK

Due to limited resources of sensor nodes, important design goals for WSNs comprise: (a) minimizing the total energy consumption within the network (b) minimizing the overhead of control messages, (c) achieving fault-tolerance, and (d) balancing energy dissipation among the sensor nodes to avoid disconnected networks.

Some of the position and non position-based routing protocols are discussed as follows:

### 2.1 Position Based Protocols

Position-based routing protocols use location information for routing decisions. To facilitate in getting the location information, a node can either be equipped with a low power GPS module, or it may use distributed localization schemes, which are based on received signal strength indicator, time of arrival, and manual registration [13, 14, 15]. In position based protocols, it is assumed that each node knows position of the destination and its neighbours, because the locations of nodes can be used to identify their network connectivity [16, 17]. Position of neighbours is usually identified via periodic hello messages. Moreover, position-based protocols can substantially reduce the communication overhead that is caused by flooding. However, getting location information of a node and its neighbours is a costly operation in terms of message transfer overhead and energy; particularly in mobile environments, where the nodes update their position more frequently.

*Geographical Energy Aware Routing* (GEAR) [17] is an energy aware geographic protocol that is designed to increase the network lifetime of WSNs. GEAR uses energy aware metrics for neighbour selection, and balances the energy consumption among neighbours by maintaining a cost function for each neighbour. The cost function facilitates in finding the cost for reaching a neighbour that is based on the location and required energy.

*Geographical Adaptive Fidelity* (GAF) [18] also relies on the location information of the nodes. GAF extends the network lifetime by reducing the energy consumption for which it builds a geographical grid that consists of cells. Each cell contains multiple nodes, but only a single node is active at a time. Meanwhile, the remaining nodes of the cell can switch to sleep state to conserve energy [16].

*Distance Routing Effect Algorithm for Mobility* (DREAM) [19] is a position based routing scheme that is designed for mobile applications. DREAM is based on a directional forwarding approach that floods the data in a particular direction (using a certain angle) towards the sink node. The drawback of DREAM is that it maintains a routing table to store information of the network nodes. Therefore, for very large networks, DREAM maintains large routing tables, which raise scalability issues.

*Adaptive Face Routing* (AFR) [20] is a distributed geographic ad hoc routing protocol. It is based on Euclidean planar graphs, in which the nodes and edges of a plane are partitioned into regions called faces. AFR uses face routing to traverse the faces in a restricted way, and avoids exploring complete boundary of the faces that lie between the source and destination. This restriction is based on the size of the ellipse area which depends on the path length. However, if the face routing fails to reach the destination, AFR falls back and repeats the process using eclipse of double size. In [21], authors propose a new version of the AFR named *Other Adaptive Face Routing* (OAFR) with an extension that it selects the boundary points that are close to the destination.

*Greedy Perimeter Stateless Routing* (GPSR) [22] uses a hybrid mechanism that is based on greedy forwarding and face routing. It embeds the location of the destination into a packet and forwards it towards the destination using greedy forwarding approach. When the greedy forwarding fails and reaches a local maximum problem, GPSR uses face routing to route around dead-ends until the packet reaches a node that is close to the destination.

Greedy Other Adaptive Face Routing (GOAFR) [21] is a distributed geographic routing protocol that uses greedy forwarding and *Other Adaptive Face Routing*. It forward the packets towards the destination using greedy forwarding approach. However, unlike face routing in GPSR, it uses OAFR for recovery.

There are some other protocols that are designed for mobile *ad hoc networks*, but they cannot be applied to WSNs directly, because they are not optimized for resource constrained environment of WSNs and are not energy aware. Few examples of such protocols include, *Ad hoc On-Demand Distance Vector* (AODV) [23], *Location Aided Routing* (LAR) [24], and *On Demand Multi path Distance Vector Routing* in *Ad Hoc Networks* (AOMDV) [4]. Moreover, the number of nodes in WSNs is comparatively very large as compared to MANETs [25]. In addition, the capabilities of sensor nodes are limited as compared to the MANET nodes. Therefore, it is not practical to use MANET protocols for the WSNs.

Hierarchical routing is widely investigated for ad hoc networks [8, 26-30] due to their attractive characteristics, such as energy efficiency and scalability [31]. In hierarchical routing protocols, the defined sensing filed is divided into regions called clusters. Each cluster contain multiple nodes, out of which, one node is designated as a cluster head. The remaining nodes associate with the cluster head and send their data. The cluster head is responsible for collection and aggregation of data from the associated nodes, and sending it to the base station.

The selected position-based protocols that are used for performance comparison are MAR [32] and GRC. Both of these protocols are specifically designed for mobile sensor networks, and they inherit the hierarchal routing characteristics, such as scalability and energy efficiency.

*2.1.1 Mobility Aware Routing (MAR)*

MAR is a hierarchal position-based routing protocol in which the sensing field is divided into a geographic grid, and cluster heads are selected on the basis of mobility factor of the nodes. Mobility factor refers to the number of times a node has moved from one zone to another. The objective of selecting cluster heads on the basis of mobility factor is to select a node as a cluster head that has minimum mobility. Therefore, during the cluster head selection process, a node having the smallest mobility factor is selected as a cluster head. This improves connectivity of the cluster head with the associated nodes. However, the major issue with this protocol is that it does not consider node energy in the cluster head selection process. Therefore, it is not an energy aware routing protocol. Moreover, it does not make full use of location information of the nodes, which results as an increase in the packet loss. Furthermore, it also incurs packet loss during the inter-cluster communication, as the cluster heads move out of the transmission range of each other.

*2.1.2 Geographic Robust Clustering (GRC)*

GRC is an energy aware protocol that uses location information during the selection of the cluster heads. In addition, it uses a recovery strategy for reducing packet loss during the inter-cluster communication phase. During the cluster head selection phase, a node is selected as a cluster head that is either at the center or close to the center, and has high residual energy. Each node calculates the weight on the basis of its residual energy and center-ness. The equation used for weight calculation is presented as follows:

$$weight = w_1 \times E - w_2 \times C \qquad (1)$$

Where, $\sum_{i=1}^{2} w_i = 1$ and $0 < w_2 < w_1$.

$E$ is the remaining energy of the sensor node, and $C$ represents the center-ness of the node that is calculated as follows:

Let (x, y) be the $x$ and $y$ coordinates of a mobile node '$m_i$' in a 2D plane, where $1 \leq i \leq n$. Also, let $(x_c, y_c)$ be the approximate center point of the region in which '$m_i$' is located. Using the coordinates and center point of the node, the value of C can be calculated as:

$$C = |x_c - x| + |x_y - y| \qquad (2)$$

In addition to the aforementioned objectives, the selection of cluster heads using this technique serves three purposes.

i. Cluster head is located either at the center or close to the center of a zone having transmission range of 'r'. Therefore, size of each cell (zone) $r/\sqrt{2} \times r/\sqrt{2}$ provides better coverage for mobile nodes during the intra-cluster communication phase.
ii. During inter-cluster communication, a cluster head may incur less packet loss, because it is more likely that both the cluster heads are within the transmission range of each other.
iii. If both cluster heads are not within transmission range of each other, then recovery strategy can be applied to avoid packet loss.

After selection of the cluster heads, normal nodes send their information to their respective cluster heads, followed by inter-cluster communication phase. Recovery strategy is applied during inter-cluster communication to reduce packet loss.

**2.2 Non Position-based Protocols**

In non position-based routing protocols, all the routing decisions are made without any location information. Therefore, it is not mandatory for the nodes to get their location information, or keep location information of their neighbours. Few well known non position-based protocols are discussed as follows.

In directed diffusion [33], data is named as attribute-value pairs. Therefore, when the sink wants to collect a specific data, the sink broadcasts interest to the nodes. The nodes save this information, and whenever data is validated against the sink's interest, it is sent to the sink. The major benefit of directed diffusion is that it provides data aggregation with optimal number of transmissions.

*Low Energy Adaptive Clustering Hierarchy* (LEACH) [8] is one of the early cluster based hierarchical non location-based routing protocol. It is designed for static networks having a fixed base station. The cluster heads are changed over a period of time, and are selected in such a way that energy utilization is evenly distributed among the nodes. LEACH is based on the assumption that each node can reach the sink node directly. This is not a realistic assumption, because sensor nodes have limited transmission range and it is not feasible for all sensor nodes to reach the sink node directly.

SPIN [34] is a data-centric protocol that avoids passing redundant data and saves energy by performing negotiation among the nodes. To achieve this, SPIN protocol names the data (meta-data), and distributes the meta-data in the network through advertising. However, nodes advertise the data to only interested neighbors. In SPIN, there is no specific format for meta-data definition as it varies from application to application.

In [33], the authors' present hybrid *energy efficient distributed* (HEED) clustering approach. HEED extends the network lifetime by selecting cluster heads on the basis of node's residual energy. To further enhance the performance, it considers intra-cluster communication cost as a secondary clustering parameter. HEED outperforms many of the clustering protocols. However, it has some pitfalls which include complex cluster head selection process (based on probabilistic methods) and support for static networks only.

The non position-based protocols that we have selected for the performance comparison are DECA [35] and DEMC [9]. Both of these protocols are hierarchal clustering based and are specifically designed for mobile sensor networks.

### 2.2.1 Distributed Efficient Clustering Approach (DECA)

In DECA [35], each node has a weight that is computed on the basis of node residual energy, connectivity, and node identifier. The positive point about this protocol is that each node transmits only one message, rather than going through rounds of iterations of probabilistic message announcements (as in LEACH and HEED). The process of message exchange during the protocol operation consumes more energy as compared to sensing and computation [28]. Therefore, by reducing the number of messages during formation of clusters lead to high energy efficiency [36]. To do so, each node maintains a neighbouring list (table) that is updated through periodic *Hello* messages. Therefore, upon receiving the clustering messages, a node decides whether it should select a cluster head or become a cluster head itself. Simulation results show that DECA has outperformed many clustering protocols including HEED. However, DECA has some pitfalls. For instance, it uses periodic hello messages for table maintenance that requires a considerable amount of energy and processing. The transmission frequency periodic hello messages increase with the increase in mobility. Consequently, it is not a good approach to maintain a table in highly mobile/dynamic environments. Nevertheless, cases may occur, where the neighbouring cluster heads move out of the transmission range of each other and incur packet loss during the inter-cluster communication phase.

### 2.2.2 Distributed Efficient Multi-hop Clustering protocol (DEMC)

DEMC focuses optimum cluster head selection to maximize network coverage. In this protocol, each node sends less than one message during the clustering phase. The difference between DEMC and DECA is that in DEMC nodes do not send periodic hello messages. Therefore, DEMC does not maintain a complete neighbour list. By doing so, DEMC reduces the number of transmissions (*Hello* messages) per node, resulting in better energy efficiency and low processing overhead. Moreover, both protocols vary in weight calculation process for the cluster head selection, and the nodes response mechanism after receiving the cluster head announcement message. During the cluster head selection process of DEMC, each node calculates the weight based on its residual energy and unique node identifier. The weight calculation equation of DEMC is mentioned as follows:

$$weight = w_1 \times E + w_2 \times I \qquad (3)$$

Where $\sum_{i=1}^{2} w_i = 1$ and $0 < w_2 < w_1$.

E is the residual energy of sensor node, and *I* is the node identifier that is used to break the tie in case two nodes have the same residual energy. For cluster head selection, each node sets a timer based on its calculated weight. Therefore, the timer of nodes that have more weight expires first, and they broadcast cluster head announcements. Consequently, when a node receives a cluster head announcement having a weight that is greater than its own weight, then for that specific round the receiving node will not send its cluster head announcement. This technique makes DEMC more energy efficient compared to DECA, as it requires less communication. Moreover, DEMC uses a recovery strategy during the inter-cluster communication phase and achieves robustness against packet loss that occurs due to node mobility.

## 3. COMPARISON STRATEGY

A variety of routing protocols are available for sensor networks. Among these, clustering based protocols are proven to be energy efficient and scalable. However, mobile sensor networks pose some different challenges for routing protocols due to their dynamic topology. The mobility and changing topology causes packet loss in two ways. Firstly, during the intra-cluster communication phase, when ordinary nodes are unable to send their information to the cluster head. Secondly, during the inter-cluster communication phase, when a cluster head cannot send the aggregated data of the whole round to another cluster head. In position based protocols, the first case can be handled by using the location information of nodes during the selection of cluster heads, as proposed in GRC. Alternatively, in non position-based protocols, this case is difficult to handle due to unavailability of location information. Consequently, large packet loss occurs in non location based protocols.

Considering the aforesaid challenge, there is a need for a mechanism to reduce the packet loss during inter-cluster communication. In [37], authors propose to increase the transmission range of nodes by a factor of $(1+ \sqrt{5})$ to ensure guaranteed connectivity. However, increase in the transmission range by such factor will also increase the energy consumption substantially that may result as a dramatic decline in the network lifetime. Alternatively, a recovery strategy can be used during the inter-cluster communication to avoid the packet loss.

Currently, there are two main approaches for packet recovery, namely hop-by-hop and end-to-end [38]. Hop-by-hop recovery is more energy efficient compared to end-to-end approach due to its shorter retransmission range. Therefore, in our experiments we use hop-by-hop recovery strategy. In DEMC and GRC a recovery strategy is applied between two cluster heads during inter-cluster communication to increase robustness of the routing protocol in terms of connectivity and resilience against the packet loss.

In this study, we select total of 6 protocols from the categories of position and non position-based protocols. For non position-based protocols, we select DECA, DEMC, and DEMC with recovery. Similarly, for position-based protocols, we select MAR, GRC, and GRC with recovery.

### 3.1 Performance Metrics

The following performance metrics are used for evaluating the aforementioned protocols.

*1) Percentage of Packet loss*

The percentage of packet loss is calculated by dividing the total number of lost packets by total number of transmitted packets. This performance metric is used to measure the robustness of protocol with respect to different node speeds. A protocol that incurs less packet loss as compared to other protocols is considered more robust against packet loss.

*2) Packet delivery ratio*

Packet delivery ratio is the ratio of the number of packets that are successfully delivered to the destination to the total number of packets that are sent by the source. This metric provides an indication of the ability of a protocol to deliver packets to the respective destination. Hence, high packet delivery ratio indicates better protocol performance.

*3) Network lifetime*

Network lifetime is defined as the number of round when the first node dies in the network. Network lifetime depends on the average energy consumption of a node per round. Therefore, the protocols that exhibit low average energy consumption per round provides longer network lifetime.

*4) Average number of control packets*

This parameter presents the average number of control messages that are required for protocol operation. As transmitting and receiving of packets consumes considerable amount of energy, it has high impact on the network lifetime.

### 3.2 Network model

A mobile wireless sensor network is modelled as a set of $V$ nodes that are interconnected by a set of full-duplex $E$ communication links. A unique identifier is used for the identification of each node. In geographic routing protocols, the nodes identify their position by either using GPS or by using some other localization mechanism. Moreover, two nodes are considered as neighbours if they have a link between them and they are within the transmission range of each other. Furthermore, nodes may move at any time, without any notice and may change the topology.

The problem of clustering can be defined as selection of cluster heads from a set of $V$ nodes in such a way that the cluster heads cover the whole network. In location based clustering protocols, each node $v$ in a set $V$ located in zone $z_i$ (where '$i$' is the number of zones), $v$ must be a cluster head or associated with a cluster head. Alternatively, in non location based clustering protocols, each node $v$ in a set $V$ must be a cluster head or associated with only one cluster head. After cluster head selection, every normal node in the cluster must be able to directly communicate with the cluster head with whom it is associated. The clustering protocol must be completely distributed, where no central control authority is required, and each node makes its decision independently based on the location information.

### 4. RESULTS & DISCUSSION

The simulations are performed in OMNET++ based simulation framework called INET [39]. INET framework supports multiple mobility models [40] and is well suited for simulation of wireless sensor networks. In this experiment, we use disk graph model for the communication links, which means that if node 'X' can reach node 'Y', then node 'Y' can also reach node 'X'. Moreover, the nodes use MAC (implemented using CSMA-CA scheme) and physical layer of 802.11. However, more efficient MAC schemes, such as [41] can also be used. Furthermore, we use an energy consumption model to monitor the network lifetime of selected protocols.

According to this model [42], the energy consumed in transmission of a k bit message over a distance d is calculated as:

$$E_{Tx}(k,d) = E_{Tx}-elec(k) + E_{Tx}-amp(k,d)$$
$$E_{Tx}(k,d) = E_{elec} \times k + E_{amp} \times k \times d^2 \qquad (4)$$

Moreover, the energy consumed in receiving a packet is given by:

$$E_{Rx}(k) = E_{Rx}-elec(k)$$
$$E_{Rx}(k) = E_{elec} \times k \qquad (5)$$

Where $E_{Tx}(k,d)$ is the energy required to transmit a 'k' bit message over a distance of d meters and $E_{Rx}(k)$ is the energy required to receive a k bit message. Moreover, $E_{elec}$ is the energy consumed for running the transceiver circuitry, and $E_{amp}$ is the energy consumed by the amplifier to achieve an acceptable Signal to Noise Ratio (SNR).

Initially, 100 nodes are randomly distributed in the network field having dimensions of 1000m × 1000m. In addition, we use Mass Mobility model for simulations, which is a variant of random waypoint mobility model, and is provided by the INET framework. This mobility model is designed to model nodes movement during which the nodes have a mass and momentum. Therefore, the nodes do not start, stop, or turn abruptly. Table 1 shows the simulation parameters that are used for simulations.

Table 1. Simulation Parameters

| Type | Parameter | Value |
| --- | --- | --- |
| Network | Area | 1000×1000 |
| | Node energy at startup | 3 J/battery |
| | Node deployment | Random |
| | Number of zones (for position-based protocols) | 16 |
| Application | Data packet size | 100 bytes |
| | Broadcast packet size | 25 bytes |
| | Packet header size | 25 bytes |
| Radio model | $E_{elec}$ | 50nJ/bit |
| | $E_{amp}$ | 0.0013 pJ/bit/m$^4$ |

Figure 1 presents packet delivery ratio with respect to different number of nodes having speed of 5 meter/second. The result shows that location based protocols attain high packet delivery ratio compared to non position based protocols. In both the categories, GRC with recovery and DEMC with recovery supersede their respective contenders. The reason for high packet delivery ratio of both the protocols is that they use recovery strategies during the intercluster communication phase. Besides, recovery strategy of GRC makes better use of location information during the cluster head selection. Consequently, the packet loss during the intracluster communication phase is low, which leads to increased packet delivery ratio. However, the location based routing itself reduces packet loss by a huge margin. Therefore, case may occur, where a protocol with recovery strategy may provide similar packet delivery ratio as without a recovery strategy.

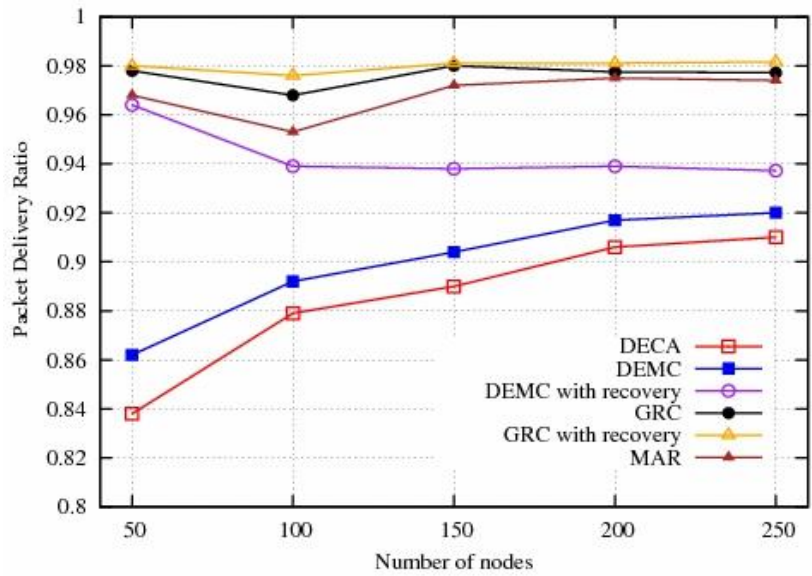

Figure 1: Packet delivery ratio with respect to different number of nodes

Figure 2 presents packet delivery ratio with respect to different node speeds. The result shows that the packet delivery ratio of the routing protocols decreases with the increase in mobility. However, the position-based routing protocols provide better delivery ratio as compared to non position based routing protocols. It is evident from Figure 3 that GRC with recovery provides highest delivery ratio among the selected protocols. Moreover, in non position-based protocols, DEMC with recovery provides the best connectivity. It can be concluded from the result that there is a need for a recovery strategy in cluster based routing protocols, as it helps to improve the packet delivery ratio during mobility.

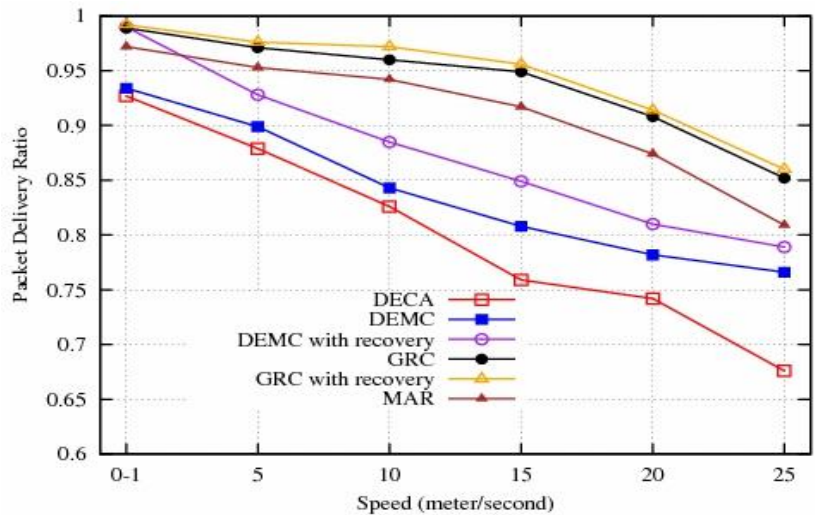

Figure 2: Packet delivery ratio with respect to different node speeds

Figure 3 shows percentage of packets lost with respect to different node speeds. The result shows that packet loss increases with the increase in mobility. It is observed during the simulations that the majority of packets are lost during the intra-cluster communications when normal nodes send information to their respective cluster heads. However, due to the node mobility, either the cluster head moves out of the transmission range of a normal node, or a normal node moves out of the transmission range of a cluster head. It is evident from the figure that position-based routing protocols perform well in terms of packet loss under variably mobility. Moreover, using recovery strategy during the inter-cluster communication phase reduces the packet loss. In non position-based routing protocols, DEMC with recovery incurs the least packet loss as compared to DECA and DEMC without recovery. Similarly, in position-based protocols, GRC with recovery incurs least packet loss as compared to MAR and GRC. It is observed that usage of recovery strategy can minimize 75% to 90% packet loss during the inter-cluster communication.

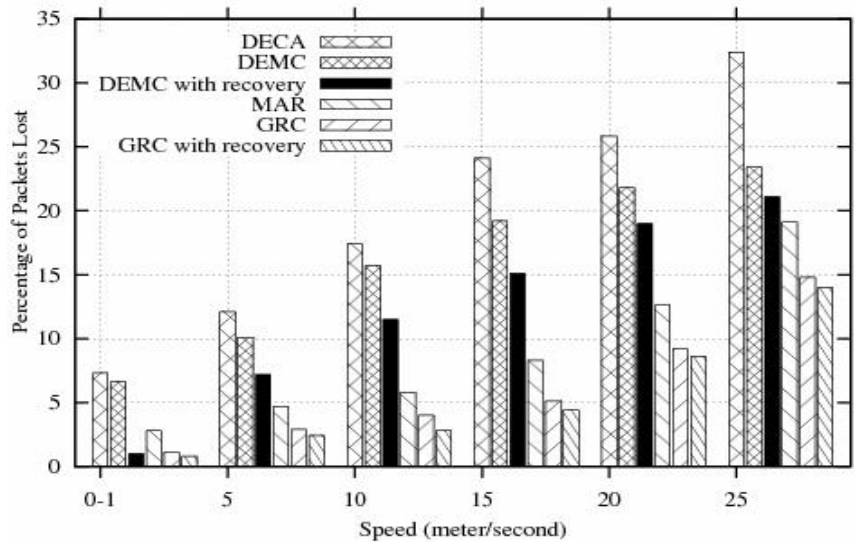

Figure 3: Percentage of packets lost with respect to different node speeds

Figure 4 shows percentage of packets lost with respect to different number of nodes. The result shows that position-based protocols perform well as compared to non position-based protocols. The primary reason behind this is that the location based routing protocols make better use of the location information while performing the clustering process. GRC with recovery performs well as compared to other protocols, because it selects a cluster head on the basis of center-ness. This factor ensures that a node is selected as cluster head that is located either at the center or near to the center of a zone. Selecting such cluster heads provide better coverage and thus reduces packet loss during the inter-cluster communication phase. Moreover, during intercluster communication, a recovery strategy is applied to avoid packet loss. As a result, GRC with recovery provides better packet delivery ratio compared to respective contender protocols.

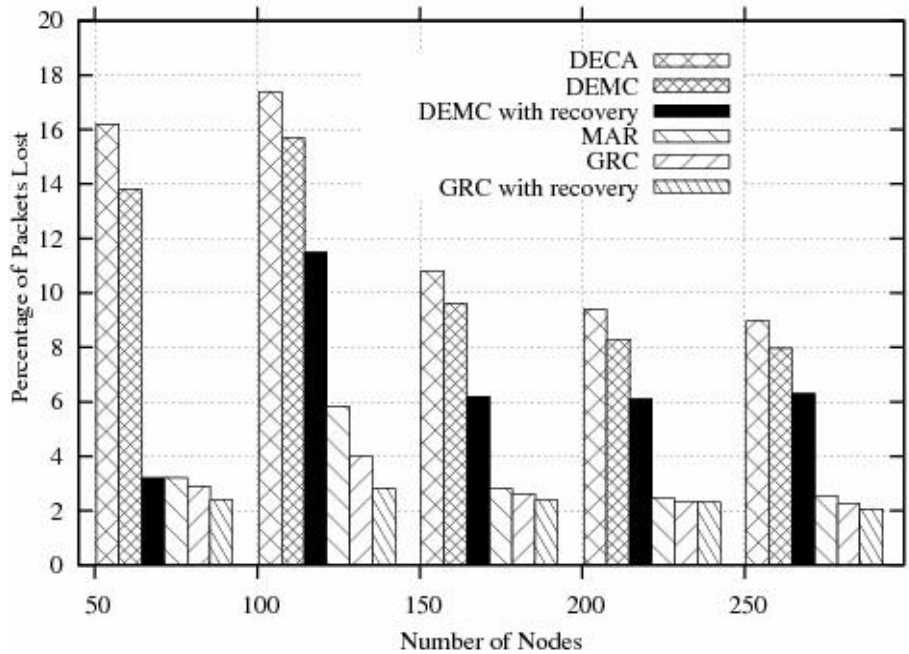

Figure 4: Percentage of packets lost with respect to different number of nodes

Figure 5 shows the average number of packets sent per node during the cluster head selection phase. In DECA, the average number of packets sent per node is one (at all times), whereas for other protocols (DEMC, MAR, and GRC), the number of packets decreases with the increase in the number of nodes. It is because, when a node receives a cluster head announcement that has more weight than its own, then for that specific round the node will not send its announcement. To do so, DEMC uses aforesaid timers for sending cluster head announcements. By using this technique, DEMC sends least average number of packets during the cluster head selection phase, which not only reduces the computational overhead but also helps to achieve energy efficiency.

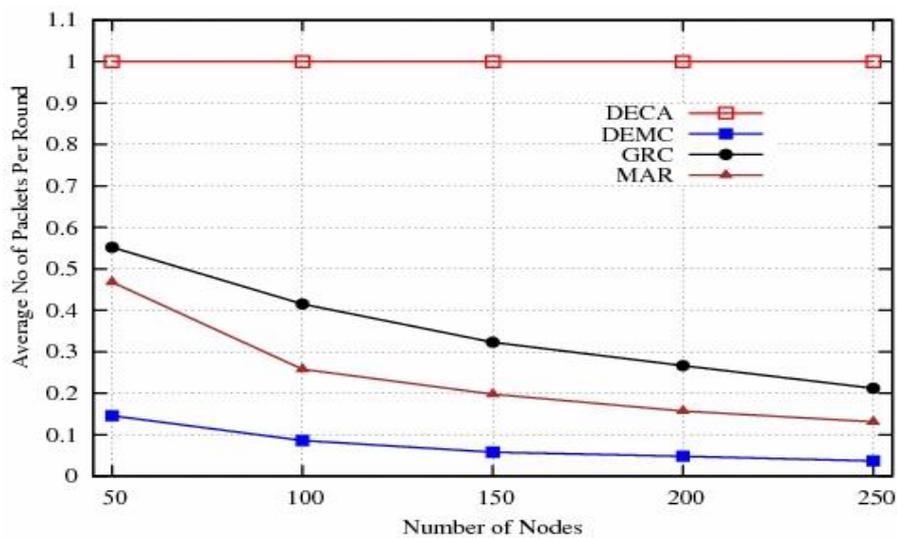

Figure 5: Average number of packets sent during cluster head selection

Figure 6 shows the network lifetime with respect to different node densities. The network lifetime can be defined as a round, in which the first node dies, or the number of dead nodes reaches a pre-defined percentage, or the number of dead nodes reaches a level where the routing to the destination is no more possible. However, as we are experimenting with the clustering based protocols, in which the energy is evenly distributed throughout the mobile network, we consider the first scenario for the definition of the network lifetime. It is because, when the first node dies, the number of dead nodes increases in the later rounds, and within 5-10 rounds the whole network becomes nonoperational. According to results, non position-based routing protocols outperform position-based protocols in terms of network lifetime. The primary reason for this behaviour is that location-based protocols consume energy in terms of localization services. Moreover, the number of control messages plays a vital role in the network lifetime. As DEMC is a non position-based based protocol and uses less control messages, it provides highest network lifetime among selected protocols, and it further increases with the increase in the number of nodes.

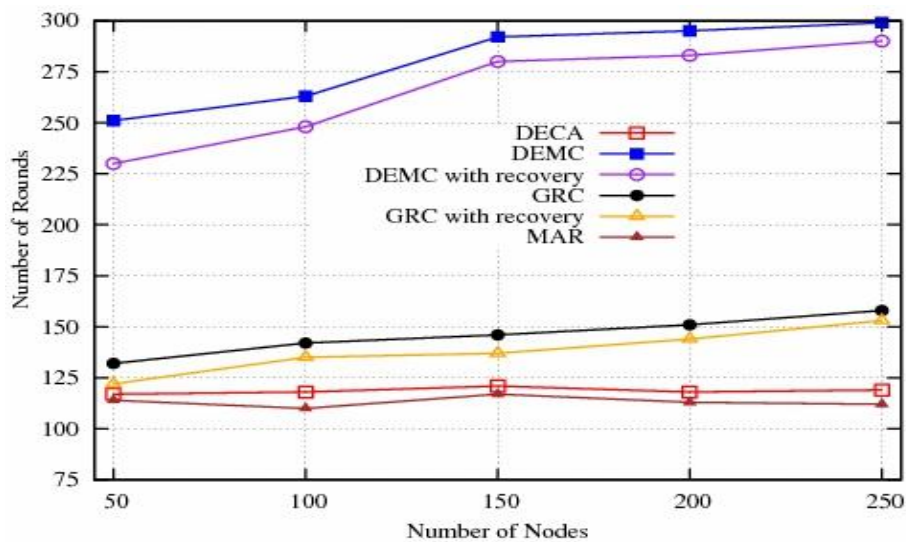

Figure 6: Network lifetime with respect to different number of nodes

Figure 7 presents the network lifetime with respect to different node speeds. The result shows that the network lifetime of non location-based routing protocols (DECA, DEMC, and DEMC with recovery) is high as compared to position-based protocols. The reason behind this is that the nodes position-update frequency increases with the increase in mobility, which is an energy consuming process. Alternatively, in non position-based protocols (DECA and DEMC), the speed of nodes does not have a major impact on the network lifetime. However, the recovery strategy factor is worth consideration, as it is an energy consuming process. As shown in Figure 7, the need for recovery strategy increases with the increase in mobility and affects the overall network lifetime. This behaviour is shown by the curves of DEMC with recovery and GRC with recovery protocols.

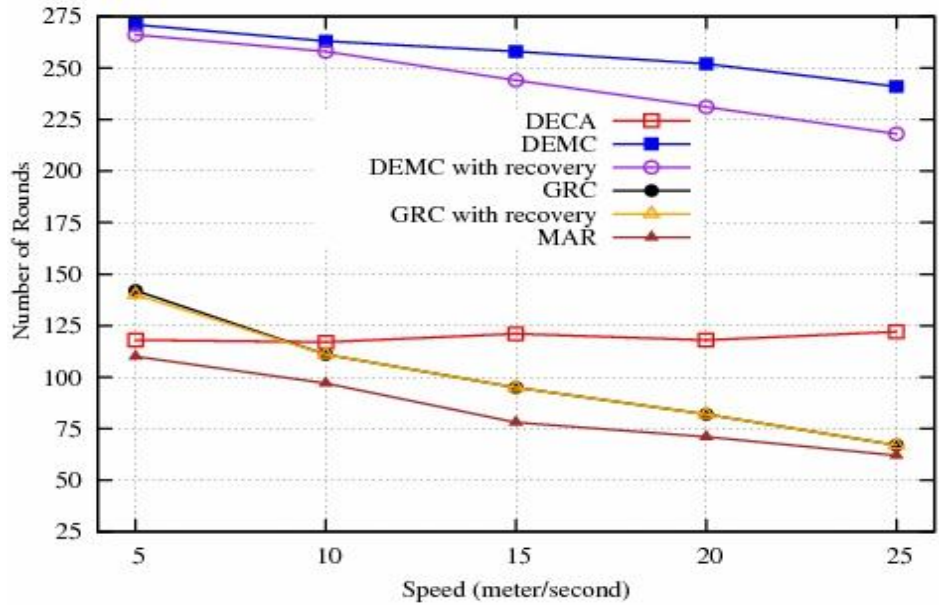

Figure 7: Network lifetime with respect to different node speeds

## 5. CONCLUSIONS & FUTURE WORK

In this paper we present performance comparison of position and non position-based hierarchal clustering protocols for mobile sensor networks. Through extensive simulations and analysis, we conclude that mobile sensor networks require a recovery strategy during inter-cluster communication, because during high mobility the cluster heads goes out of the transmission range of each other. Moreover, position-based routing protocols incur less packet loss and provide high packet delivery ratio as compared to non position-based protocols, due to optimal cluster head selection. However, position-based protocols rely on location information for their operation. Therefore, they require localization mechanisms, which are considered costly in terms of energy consumption. Consequently, such protocols may deliver low network lifetime.

Considering the sensor network application and trade off between performance parameters (packet loss, packet delivery ratio, and network lifetime), the routing protocols must be selected accordingly. If the desired objective is to increase the network lifetime under moderate node mobility, then non position-based protocols, such as DEMC with recovery is the best choice. DEMC is suitable for multiple applications, such as monitoring growth of plants, soil moisture, or habitat monitoring. Alternatively, if an application requires high packet delivery ratio and there are no severe energy constraints, then usage of position-based protocols is recommended. GRC is suitable time critical applications, such as military surveillance, fire, seismic, and flood detection.

For future work, we aim to implement cross layer design for mobile sensor networks to achieve enhanced energy efficiency and robustness against packet loss. Moreover, we intend to evaluate the performance of the aforementioned routing protocols using different mobility models [43, 44].

**References**

[1]  Tashtarian, F., Haghighat, A. T., Honary, M. T., & Shokrzadeh, H. (2007, September). A new energy-efficient clustering algorithm for wireless sensor networks. In *15th International Conference on Software, Telecommunications and Computer Networks, 2007 (SoftCOM)* (pp. 1-6). IEEE.

[2] Wang, Q., & Yang, W. (2007, June). Energy consumption model for power management in wireless sensor networks. In *4th Annual IEEE Communications Society Conference on Sensor, Mesh and Ad Hoc Communications and Networks (SECON'07)* (pp. 142-151). IEEE.

[3] Khan, A. R., Bilal, S. M., & Othman, M. (2012, November). A performance comparison of open source network simulators for wireless networks. In *International Conference on Control System, Computing and Engineering (ICCSCE)* (pp. 34-38). IEEE.

[4] Marina, M. K., & Das, S. R. (2001, November). On-demand multipath distance vector routing in ad hoc networks. In *Ninth International Conference on Network Protocols* (pp. 14-23). IEEE.

[5] Yang, T., Tetsuya, O., Gjergji, M., Leonard, B., et al (2012). Energy-saving in wireless sensor networks considering mobile sensor nodes, single and multi events. International Journal of Computer Systems Science and Engineering 27(5).

[6] Khan, A. R., Othman, M., Madani, S., & Khan, S (2013). A Survey of Mobile Cloud Computing Application Models. IEEE Communications Surveys and Tutorials PP(99) 1-21.

[7] Anastasi, G., Conti, M., Di Francesco, M., & Passarella, A. (2009). Energy conservation in wireless sensor networks: A survey. *Ad Hoc Networks*, *7*(3), 537-568.

[8] Heinzelman, W. B., Chandrakasan, A. P., & Balakrishnan, H. (2002). An application-specific protocol architecture for wireless microsensor networks. In *IEEE Transactions on Wireless Communications, 1*(4), 660-670.

[9] Ali, S., & Madani, S. (2011). Distributed efficient multi hop clustering protocol for mobile sensor networks. *Int. Arab J. Inf. Technol.*, *8*(3), 302-309.

[10] Fang-Yie, L., Fuu-Cheng, J., Chih-Cheng, L., Sen-Tarng L et al (2012). A rate-allocation based multi-path control scheme for event-driven wireless sensor networks on constant event packet rates, International Journal of Computer Systems Science and Engineering 27(5).

[11] Tao, Y., Tetsuya, O., Gjergji, M., Leonard, B. et al (2012). Energy-saving in wireless sensor networks considering mobile sensor nodes, single and multi events, International Journal of Computer Systems Science and Engineering 27(5).

[12] Giordano, S., Stojmenovic, I., & Blazevic, L. (2003). Position based routing algorithms for ad hoc networks: A taxonomy. *Ad hoc wireless networking*, *8*, 64.

[13] Basagni, S. (1999). Distributed clustering for ad hoc networks. In *Fourth International Symposium on Parallel Architectures, Algorithms, and Networks (I-SPAN'99)* (pp. 310-315). IEEE.

[14] Stojmenovic, I. (2002). Position-based routing in ad hoc networks. In *IEEE Communications Magazine, 40*(7), 128-134.

[15] Jin, Z., Jian-Ping, Y., Si-Wang, Z., Ya-Ping, L., & Guang, L. (2009). A survey on position-based routing algorithms in wireless sensor networks. *Algorithms*, *2*(1), 158-182.

[16] Seada, K., & Helmy, A. (2004). Geographic protocols in sensor networks. *Encyclopedia of Sensors, American Scientific Publishers (ASP)*.

[17] Yu, Y., Govindan, R., & Estrin, D. (2001). *Geographical and energy aware routing: A recursive data dissemination protocol for wireless sensor networks*. Technical report ucla/csd-tr-01-0023, UCLA Computer Science Department.

[18] Xu, Y., Heidemann, J., & Estrin, D. (2001, July). Geography-informed energy conservation for ad hoc routing. In *Proceedings of the 7th annual international conference on Mobile computing and networking* (pp. 70-84). ACM.

[19] Basagni, S., Chlamtac, I., Syrotiuk, V. R., & Woodward, B. A. (1998, October). A distance routing effect algorithm for mobility (DREAM). In *Proceedings of the 4th annual ACM/IEEE international conference on Mobile computing and networking* (pp. 76-84). ACM.

[20] Kuhn, F., Wattenhofer, R., & Zollinger, A. (2002, September). Asymptotically optimal geometric mobile ad-hoc routing. In *Proceedings of the 6th international workshop on Discrete algorithms and methods for mobile computing and communications* (pp. 24-33). ACM.

[21] Kuhn, F., Wattenhofer, R., & Zollinger, A. (2003, June). Worst-case optimal and average-case efficient geometric ad-hoc routing. In *Proceedings of the 4th ACM international symposium on Mobile ad hoc networking & computing* (pp. 267-278). ACM.

[22] Karp, B., & Kung, H. T. (2000, August). GPSR: Greedy perimeter stateless routing for wireless networks. In *Proceedings of the 6th annual international conference on Mobile computing and networking* (pp. 243-254). ACM.

[23] Das, S. R., Belding-Royer, E. M., & Perkins, C. E. (2003). Ad hoc on-demand distance vector (AODV) routing.

[24] Ko, Y. B., & Vaidya, N. H. (2000). Location-Aided Routing (LAR) in mobile ad hoc networks. *Wireless Networks*, *6*(4), 307-321.

[25] Bilal, S. M., Dilber, M. N., Khan, A. R. (2013). Routing Proposals for Multipath Interdomain Routing, 15th International Multi Topic Conference (pp. 331-337). IEEE.

[26] Khan, A. R., Madani, S. A., Hayat, K., & Khan, S. U. (2012). Clustering-based power-controlled routing for mobile wireless sensor networks. *International Journal of Communication Systems*, *25*(4), 529-542.

[27] Ali, S., & Madani, S. (2009). Distributed efficient multi hop clustering protocol for mobile sensor networks. *Int. Arab J. Inf. Technol.*, *8*(3), 302-309.

[28] Patwari, N., Ash, J. N., Kyperountas, S., Hero III, A. O., Moses, R. L., & Correal, N. S. (2005). Locating the nodes: cooperative localization in wireless sensor networks. *In IEEE Signal Processing Magazine, 22*(4), 54-69.

[29] Chan, K. S., Pishro-Nik, H., & Fekri, F. (2005, March). Analysis of hierarchical algorithms for wireless sensor network routing protocols. In *Wireless Communications and Networking Conference* (pp. 1830-1835). IEEE.

[30] Lindsey, S., & Raghavendra, C. S. (2002). PEGASIS: Power-efficient gathering in sensor information systems. In *Aerospace conference proceedings* (pp. 3-1125). IEEE.

[31] Chan, K. S., Pishro-Nik, H., & Fekri, F. (2005, March). Analysis of hierarchical algorithms for wireless sensor network routing protocols. In *Wireless Communications and Networking Conference* (pp. 1830-1835). IEEE.

[32] Arboleda C, L. M., & Nasser, N. (2006, August). Cluster-based routing protocol for mobile sensor networks. In *Proceedings of the 3rd international conference on Quality of service in heterogeneous wired/wireless networks* (p. 24). ACM.

[33] Younis, O., & Fahmy, S. (2004). HEED: a hybrid, energy-efficient, distributed clustering approach for ad hoc sensor networks. *IEEE Transactions onMobile Computing, 3*(4), 366-379.

[34] Heinzelman, W. R., Kulik, J., & Balakrishnan, H. (1999, August). Adaptive protocols for information dissemination in wireless sensor networks. In *Proceedings of the 5th annual ACM/IEEE international conference on Mobile computing and networking* (pp. 174-185). ACM.

[35] Yu, M., Li, J. H., & Levy, R. (2006). Mobility resistant clustering in multi-hop wireless networks. *Journal of Networks*, *1*(1), 12-19.

[36] Huang, X., Zhai, H., & Fang, Y. (2006, October). Lightweight robust routing in mobile wireless sensor networks. In *Military Communications Conference(MILCOM)* (pp. 1-6). IEEE.

[37] Ye, F., Zhong, G., Cheng, J., Lu, S., & Zhang, L. (2003, May). PEAS: A robust energy conserving protocol for long-lived sensor networks. In *Proceedings of 23rd International Conference on Distributed Computing Systems* (pp. 28-37). IEEE.

[38] Yick, J., Mukherjee, B., & Ghosal, D. (2008). Wireless sensor network survey. *Computer networks*, *52*(12), 2292-2330.

[39] Varga, A. (2001, June). The OMNeT++ discrete event simulation system. In *Proceedings of the European Simulation Multiconference (ESM'2001)* (Vol. 9, p. 185).

[40] Drytkiewicz, W., Sroka, S., Handziski, V., Köpke, A., & Karl, H. (2003). A Mobility Framework for OMNeT+.

[41] Mahlknecht, S., & Bock, M. (2004, September). CSMA-MPS: A minimum preamble sampling MAC protocol for low power wireless sensor networks. In *Proceedings of International Workshop on Factory Communication Systems* (pp. 73-80). IEEE.

[42] Heinzelman, W. R., Chandrakasan, A., & Balakrishnan, H. (2000, January). Energy-efficient communication protocol for wireless microsensor networks. In *Proceedings of the 33rd Annual Hawaii International Conference on System Sciences, 2000* (pp. 10-pp). IEEE.

[43] Khan, A. R, Ali, S. A., Mustafa, S., Madani, S. A. (2010). Behavior of clustering based routing protocols with respect to different mobility models in mobile WSNs, *Proceedings of International Conference on Intelligence and Information Technology* (pp. 402-406).

[44] Khan, A. R., Ali, S., Mustafa, S., & Othman, M. (2012, December). Impact of Mobility Models on Clustering Based Routing Protocols in Mobile WSNs. In *Proceedings of 10th International Conference on Frontiers of Information Technology (FIT)* (pp. 366-370). IEEE.